\documentclass[aps,prl,twocolumn,superscriptaddress,floatfix]{revtex4-2}
\usepackage{makecell}
\usepackage{graphicx}% Include figure files
\usepackage{dcolumn}% Align table columns on decimal point
\usepackage{bm}% bold math
\usepackage{times}% font that prx use
\usepackage{color}
\usepackage{appendix}
\usepackage{lineno}
\usepackage{epsfig}
\usepackage{amsmath}
\usepackage{amssymb}
\usepackage{mathrsfs}
\usepackage{booktabs}
\usepackage{multirow}

\begin{document}

\title{The distinction of time-reversal-like degeneracy by electronic transport in a new compound Ce$_3$MgBi$_5$}

\author{Yi-Yan Wang}%\email{wyy@ahu.edu.cn}
\affiliation{Anhui Provincial Key Laboratory of Magnetic Functional Materials and Devices, Institutes of Physical Science and Information Technology, Anhui University, Hefei, Anhui 230601, China}
\affiliation{Key Laboratory of Materials Physics, Ministry of Education, School of Physics, Zhengzhou University, Zhengzhou 450001, China}

\author{Ping Su}
\affiliation{Anhui Provincial Key Laboratory of Magnetic Functional Materials and Devices, Institutes of Physical Science and Information Technology, Anhui University, Hefei, Anhui 230601, China}

\author{Kai-Yuan Hu}
\affiliation{Anhui Provincial Key Laboratory of Magnetic Functional Materials and Devices, Institutes of Physical Science and Information Technology, Anhui University, Hefei, Anhui 230601, China}

\author{Yi-Ran Li}
\affiliation{Anhui Provincial Key Laboratory of Magnetic Functional Materials and Devices, Institutes of Physical Science and Information Technology, Anhui University, Hefei, Anhui 230601, China}

\author{Na Li}
\affiliation{Anhui Provincial Key Laboratory of Magnetic Functional Materials and Devices, Institutes of Physical Science and Information Technology, Anhui University, Hefei, Anhui 230601, China}

\author{Ying Zhou}
\affiliation{Anhui Provincial Key Laboratory of Magnetic Functional Materials and Devices, Institutes of Physical Science and Information Technology, Anhui University, Hefei, Anhui 230601, China}

\author{Dan-Dan Wu}
\affiliation{Anhui Provincial Key Laboratory of Magnetic Functional Materials and Devices, Institutes of Physical Science and Information Technology, Anhui University, Hefei, Anhui 230601, China}

\author{Yan Sun}
\affiliation{Anhui Provincial Key Laboratory of Magnetic Functional Materials and Devices, Institutes of Physical Science and Information Technology, Anhui University, Hefei, Anhui 230601, China}

\author{Qiu-Ju Li}
\affiliation{School of Physics, Anhui University, Hefei, Anhui 230601, China}

\author{Xia Zhao}
\affiliation{School of Physical Sciences, University of Science and Technology of China, Hefei, Anhui 230026, China}

\author{Hui Liang}%\email{lianghui@ahu.edu.cn}
\affiliation{Anhui Provincial Key Laboratory of Magnetic Functional Materials and Devices, Institutes of Physical Science and Information Technology, Anhui University, Hefei, Anhui 230601, China}

\author{Xue-Feng Sun}%\email{xfsun@ahu.edu.cn}
\affiliation{Anhui Provincial Key Laboratory of Magnetic Functional Materials and Devices, Institutes of Physical Science and Information Technology, Anhui University, Hefei, Anhui 230601, China}

\date{\today}

\begin{abstract}

We report the discovery of a new compound, Ce$_3$MgBi$_5$, and reveal the hidden time-reversal-like degenerate
states within it. Ce$_3$MgBi$_5$ is an antiferromagnet with the distorted kagome lattice of Ce atoms, in which several fractional magnetization plateaus emerge with the increase of magnetic field. At the 1/2 magnetization plateau, obvious hysteresis has been observed in the magnetoresistance and Hall resistivity during the rise and fall of the magnetic field. However, hysteresis vanishes in the corresponding measurements of magnetization, indicating the existence of degenerate states with the same net magnetization but different electronic transport properties. The degenerate states can be connected by the time-reversal-like operation. In addition, by comparing with HoAgGe, it is suggested that the special crystal structure in Ce$_3$MgBi$_5$ may have a shielding effect on the time-reversal-like operation, thereby affecting the distinction of degenerate states. Our work establishes Ce$_3$MgBi$_5$ as an example of utilizing electronic transport properties to identify and distinguish hidden symmetries in frustrated magnetic systems.
\end{abstract}
\maketitle

%\section{Introduction}

\emph{Introduction.} Time-reversal symmetry is one of the important fundamental concepts in physics, playing an important role in fields such as magnetism, topological physics, and superconductivity etc\cite{ghosh2020recent,PhysRev.52.361,RevModPhys.82.3045,ando2013topological,PhysRevB.83.205101,wang2026unconventional}. In magnetic materials with broken time-reversal symmetry, two time-reversal-related equilibrium states are energetically degenerate, which can be reflected in the hysteresis of magnetization or the anomalous Hall effect (AHE)\cite{zhao2024discrete}. The net magnetization or AHE usually have opposite signs but the same size in these two states. In some non-collinear antiferromagnets, even if the net magnetization disappears, the AHE can still distinguish the breaking of time-reversal symmetry, where the AHE has the same size and is driven by the nonvanishing Berry curvature\cite{nakatsuji2015large,PhysRevLett.112.017205,PhysRevB.100.045109,nayak2016large}.

Geometrically frustrated magnetic systems provide a platform to explore exotic physical phenomena. Recently, Zhao \emph{et al.} reveal the hidden time-reversal-like degenerate states in the kagome spin ice material HoAgGe through electronic transport measurements\cite{zhao2024discrete,zhao2020realization}. In HoAgGe, the Hall resistivity and magnetoresistance (MR) dependent on the magnetic field exhibit significant hysteresis, which is almost disappears in the magnetic field dependent magnetization. This observation indicates that the two related states have different sizes of AHE and MR, but have nearly the same net magnetization and energy\cite{zhao2024discrete}. In fact, hysteresis also occurs in magnetic field dependent thermal conductivity\cite{PhysRevB.106.014416}. In the crystal structure of HoAgGe, there exists the distorted kagome lattice, in which the magnetic moments of Ho atoms form clockwise and counterclockwise toroidal chiral structures. The two degenerate toroidal moment states possess opposite chirality and Berry curvature. After the chirality of the toroidal structures is reversed by the increasing magnetic field, the distribution of the corresponding degenerate states will change. The chirality reversal of the toroidal structures is related to a time-reversal-like operation, which changes the values of AHE and MR but leaves the net magnetization unchanged. However, these changes will be retained during the demagnetization process due to the influence of the distorted kagome lattice, enabling the distinction of such degenerate states to be achieved through electronic transport measurements\cite{liu2024electronic}. The work of Zhao \emph{et al.} established electronic transport as an effective means of detecting hidden states in frustrated systems.

Recently, the quasi-one-dimensional Ln$_3$MPn$_5$ (Ln = lanthanide; M = Mg, transition metal; Pn = pnictide) family materials have attracted widespread attention for the rich physical properties\cite{khoury2022class,yi2023extremely,khoury2024towards,PhysRevB.108.075157,he2025rich,PhysRevB.110.134432,PhysRevB.106.184405,duan2020highCr,duan2020highTi,hayami2022magnetic,ritter2021magnetic,li2026power,matin2017probing,PhysRevB.109.L140405,shinozaki2020magnetoelectric,shinozaki2020magnetoelectric,PhysRevMaterials.7.124406,PhysRevB.110.165106,8b7v-jc4v,li2025crystal}.
A prominent feature of the crystal structure of Ln$_3$MPn$_5$ is the distorted kagome lattice composed of rare-earth atoms, which is similar to that of HoAgGe, but with a different stacking pattern along the $c$ axis. Such structural differences may give rise to new phenomena or physics. In particular, the magnetic moments of rare-earth atoms are also always confined within the plane. Therefore, Ln$_3$MPn$_5$ materials with magnetic rare-earth atoms are promising candidates for exploring the aforementioned time-reversal-like degenerate states. Motivated by the potential interesting physical properties, we conducted exploration in Ln$_3$MPn$_5$ materials and discovered a new member, Ce$_3$MgBi$_5$.

In this Letter, we synthesized the single crystals of Ce$_3$MgBi$_5$, determined the crystal structure, and systematically studied the magnetic properties and electronic transport properties. The magnetic susceptibility exhibits antiferromagnetic (AFM) transition at 4.4 K. With the increase of magnetic field, the AFM transition can develop and evolve in the in-plane directions, but is much limited in the out-of-plane direction. Several fractional magnetization plateaus have been observed in the magnetic field dependent magnetization, indicating the complex changes of the spin state in Ce$_3$MgBi$_5$. Through electron transport measurements involving four configurations with different current and magnetic field directions, we identified the hidden time-reversal-like degenerate states at the 1/2 magnetization plateau, manifested as significant hysteresis in MR and Hall effect but vanishing hysteresis in magnetization. In addition, we qualitatively explained the reason why hysteresis occurs at the 1/3 and 2/3 plateaus in HoAgGe but at the 1/2 plateau in Ce$_3$MgBi$_5$ based on structural differences.

%\section{Results and discussion}

\begin{figure}[htbp]
\centering
\includegraphics[width=0.4\textwidth]{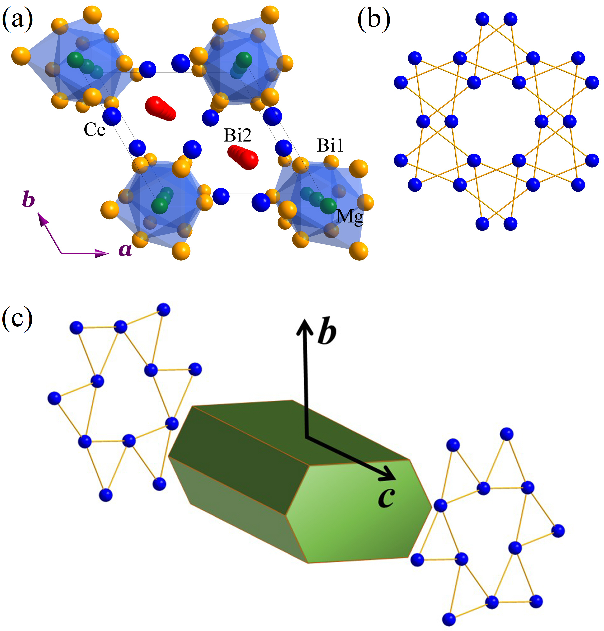}
\caption{(a) Crystal structure of Ce$_3$MgBi$_5$ viewed from the $c$-axis direction. (b) Two alternately stacked distorted
kagome lattice layers composed of Ce atoms. (c) Schematic diagram of the crystal axis directions. The long axis direction is perpendicular to the distorted kagome lattice layers.}
\end{figure}

\begin{figure*}[htbp]
\centering
\includegraphics[width=0.95\textwidth]{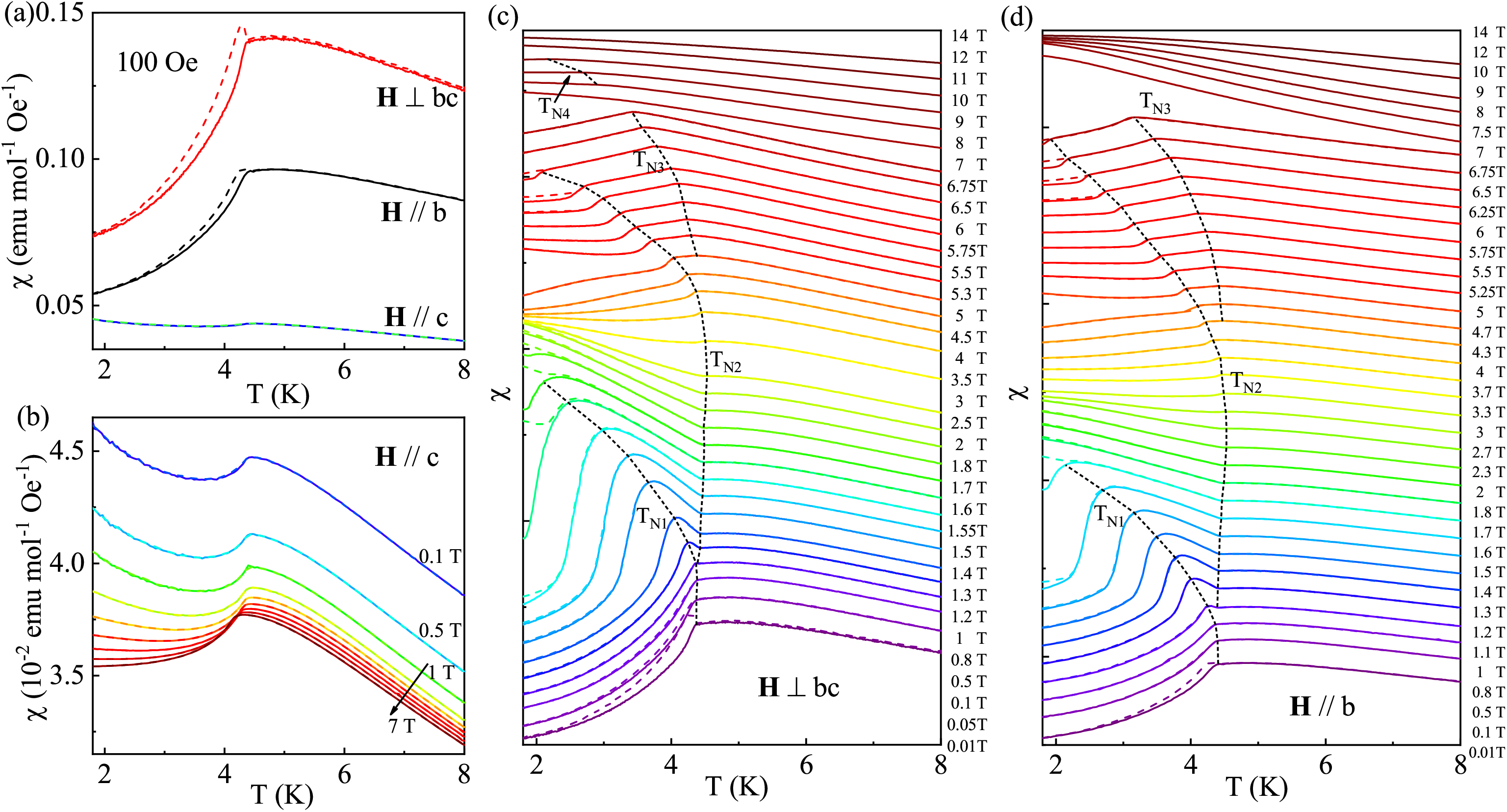}
\caption{(a) The ZFC-FC magnetic susceptibility of Ce$_3$MgBi$_5$ with applying 100 Oe magnetic field in the directions of $\textbf{H} \perp bc$, $\textbf{H} \parallel b$ and $\textbf{H} \parallel c$. (b)-(d) The ZFC-FC $\chi(T)$ curves under various magnetic fields in these three directions. In (c) and (d), to more clearly demonstrate the changes in magnetic transitions, the curves have been shifted along the vertical axis.}
\end{figure*}

\emph{Results and discussion.} Ce$_3$MgBi$_5$ crystallizes in the space group \emph{P6}$_3$/\emph{mcm} (No. 193). The crystallographic parameters are listed in the Supplemental Material\cite{SupplementalMaterial}. Figure 1(a) illustrates the determined crystal structure of Ce$_3$MgBi$_5$. Similar to other members of the Ln$_3$MPn$_5$ family, Ce$_3$MgBi$_5$ contains hypervalent 1D Bi chains and distorted kagome lattice of Ce atoms. Unlike the overlapping arrangement in HoAgGe, the adjacent distorted kagome layers in Ce$_3$MgBi$_5$ are rotated by 180$^{\circ}$, as shown in Fig. 1(b). The distorted kagome layers are located within the $ab$ plane and stacked along the $c$ axis [Fig. 1(c)].

Figure 2(a) shows the temperature dependent zero-field-cooling (ZFC) and field-cooling (FC) magnetic susceptibility of Ce$_3$MgBi$_5$ along $\textbf{H} \perp bc$, $\textbf{H} \parallel b$ and $\textbf{H} \parallel c$, all of which undergo the AFM transition at 4.4 K. The solid lines and the dashed lines represent ZFC curves and FC curves, respectively. As the temperature decreases, the FC curves within the distorted kagome lattice plane initially experience a sudden increase and then decrease at the Neel temperature, forming a difference from the ZFC curves, which may be due to the influence of magnetic fluctuation. The magnetic susceptibility exhibits significant anisotropy, with easier magnetization occurring within the plane. For the out-of-plane direction of $\textbf{H} \parallel c$, the application of a magnetic field has a minor impact on the Neel temperature [Fig. 2(b)]. However, in the in-plane directions of $\textbf{H} \perp bc$ and $\textbf{H} \parallel b$, the magnetic field can significantly affect and induce the magnetic transitions, indicating that the spin arrangement is constrained in the plane of the distorted kagome lattice. Figures 2(c) and 2(d) show the ZFC-FC magnetic susceptibility under various magnetic fields in the $\textbf{H} \perp bc$ and $\textbf{H} \parallel b$ directions, respectively. With the increase of the magnetic field, the transition at $T_{\text{N1}}$ gradually widens and shifts to lower temperature, while the transition at $T_{\text{N2}}$ begins to emerge, manifested as a sudden increase in magnetic susceptibility above $T_{\text{N1}}$. This indicates that magnetic ordering is gradually changing with temperature rather than suddenly formed, and $T_{\text{N2}}$ is only the beginning of spin changes. As the magnetic field further increases, the sudden increase in magnetic susceptibility at $T_{\text{N2}}$ evolves into a decrease, forming the AFM order. When the magnetic field exceeds 4 T, another AFM order $T_{\text{N3}}$ develops. In the range of 7-14 T, no new phase transition occurs after $T_{\text{N3}}$ is suppressed and disappears in the $\textbf{H} \parallel b$ direction. However, in the $\textbf{H} \perp bc$ direction, a new AFM order $T_{\text{N4}}$ emerges, which seems to be the reappearance of the suppressed $T_{\text{N3}}$. This difference is also reflected in the subsequent magnetic field dependent magnetization and electron transport, manifested as an additional transition in the $\textbf{H} \perp bc$ direction under high fields.

\begin{figure}[htbp]
\centering
\includegraphics[width=0.48\textwidth]{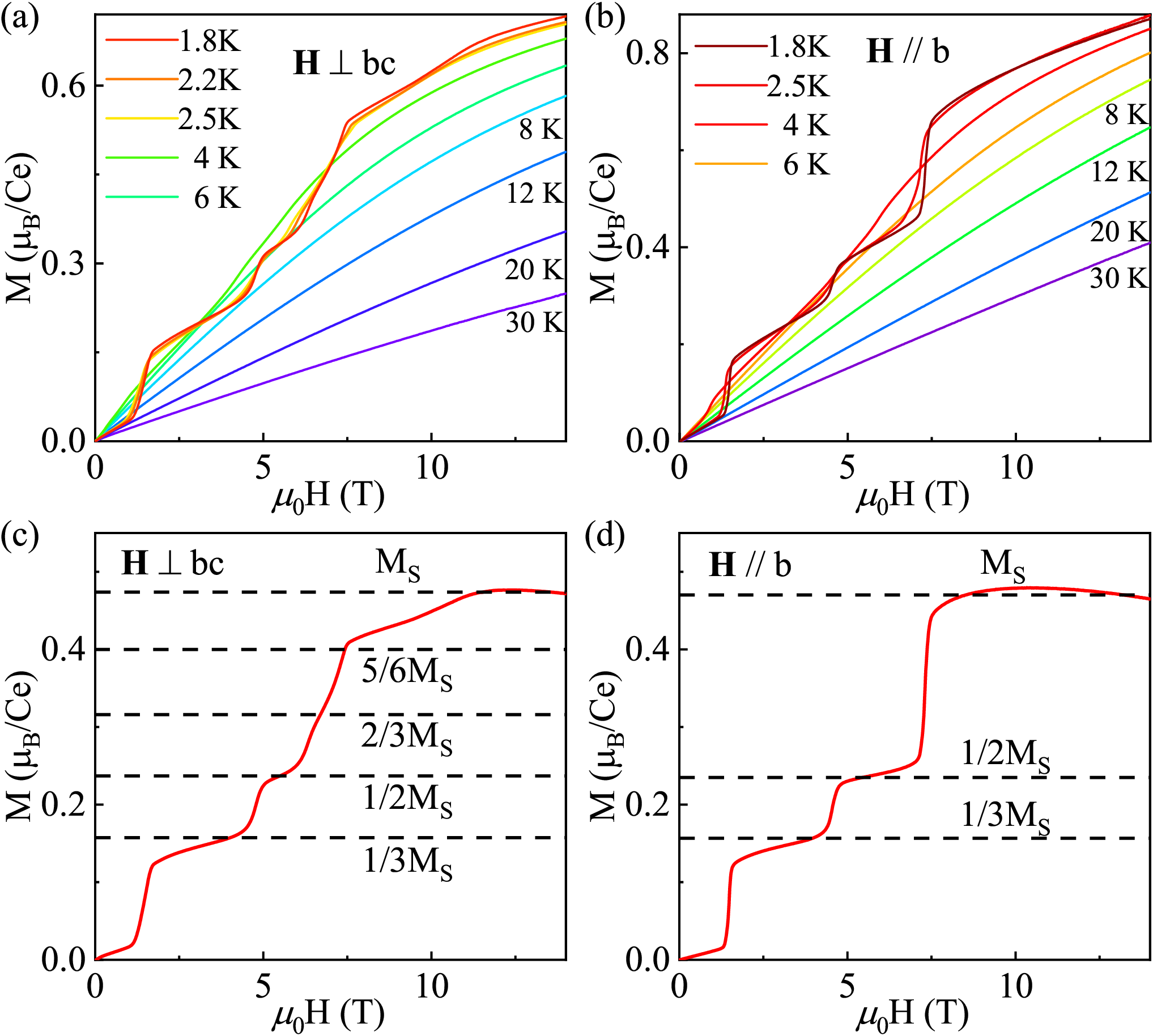}
\caption{(a) and (b) Magnetic field dependent magnetization of Ce$_3$MgBi$_5$ in the $\textbf{H} \perp bc$ and $\textbf{H} \parallel b$ directions at different temperatures. (c) and (d) Magnetization curves at 1.8 K for $\textbf{H} \perp bc$ and $\textbf{H} \parallel b$ after subtracting the Van-Vleck term.}
\end{figure}

Figures 3(a) and 3(b) plot the isothermal magnetization of Ce$_3$MgBi$_5$ as a function of magnetic field for $\textbf{H} \perp bc$ and $\textbf{H} \parallel b$, respectively. With the application of magnetic field, the in-plane magnetization of Ce$_3$MgBi$_5$ exhibits a series of metamagnetic transitions, indicating the formation of different spin states. The values of magnetization obtained at 14 T are much smaller than the saturation magnetic moment of free Ce$^{3+}$ ion (2.14 $\mu_B$), which can be mainly attributed to the shielding of Ce$^{3+}$ magnetic moment by the crystalline electric field effect\cite{matin2017probing}. No transition or anomaly was observed for $\textbf{H} \parallel c$ (Supplemental Material\cite{SupplementalMaterial}). In order to show the magnetization plateaus more clearly, a linear background usually related to the Van-Vleck paramagnetism has been subtracted from the data at 1.8 K\cite{PhysRevB.106.134426,PhysRevMaterials.4.064410}. As shown in Figs. 3(c) and 3(d), the magnetization plateaus correspond to a series of fractional magnetization phases, demonstrating the changes in spin states during the increase of magnetic field. For $\textbf{H} \parallel b$, after experiencing the 1/3M$_\text{S}$ and 1/2M$_\text{S}$ plateaus, the magnetization quickly enters the saturation state around 7.3 T [Fig. 3(d)]. However, for $\textbf{H} \perp bc$, two additional kinks can be observed at 2/3M$_\text{S}$ and 5/6M$_\text{S}$, indicating more complex changes in magnetic structure in this direction. The saturation magnetic field in $\textbf{H} \parallel b$ can only make the magnetization reach 5/6M$_\text{S}$ in $\textbf{H} \perp bc$, after which the magnetization gradually increases to the saturation value M$_\text{S}$ under larger magnetic field. In fact, the observed fractional magnetization of 1/3 (4/12), 1/2 (6/12), 2/3 (8/12), and 5/6 (10/12) can be understood based on the process of gradual spin reversal in the AFM kagome lattice.

\begin{figure*}[htbp]
\centering
\includegraphics[width=\textwidth]{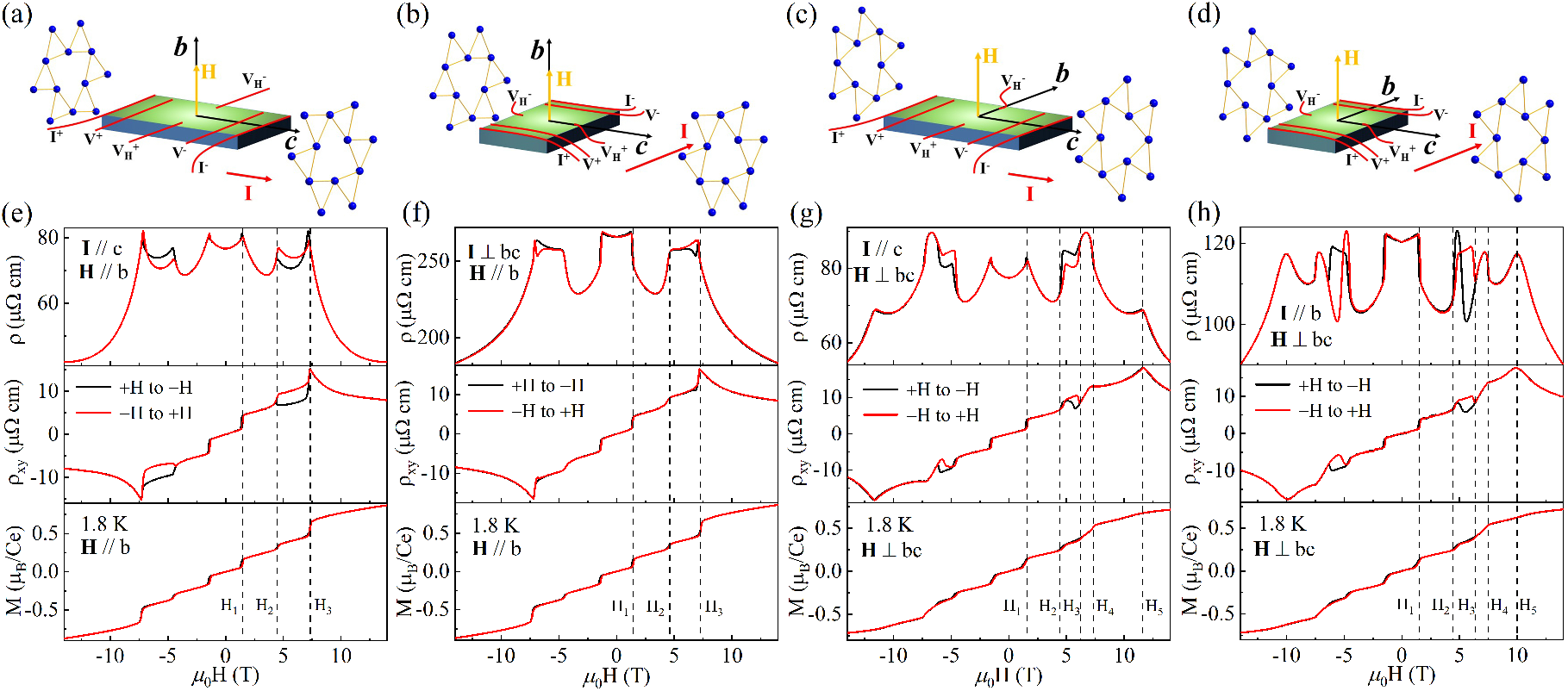}
\caption{(a)-(d) Schematic diagrams of the four electronic transport measurement configurations used. In (a) and (b), the magnetic field is along the $b$ axis, while the current flows along the $c$ axis ($\textbf{I} \parallel c$, perpendicular to the distorted kagome lattice) and perpendicular to the $bc$ direction ($\textbf{I} \perp bc$, parallel to the distorted kagome lattice), respectively. In (c) and (d), the magnetic field is perpendicular to the $bc$ direction, while the current flows along the $c$ axis ($\textbf{I} \parallel c$, perpendicular to the distorted kagome lattice) and $b$ axis ($\textbf{I} \parallel b$, parallel to the distorted kagome lattice), respectively. (e)-(h) Magnetic field dependent resistivity (upper panel), Hall resistivity (middle panel), and magnetization (lower panel) corresponding to four measurement configurations. The black lines and the red lines represent the measurement results of the magnetic field sweeping from $+$14 T to $-$14 T and from $-$14 T to $+$14 T, respectively. The temperature is fixed at 1.8 K. The data has been symmetrized or antisymmetrized to eliminate the influence of electrodes.}
\end{figure*}

Since Ce$_3$MgBi$_5$ possesses the distorted kagome lattice and exhibits the fractional magnetization plateaus, it is interesting to explore whether it has the similar time-reversal-like degenerate states to HoAgGe and distinguish them through electronic transport measurements. Four configurations are employed for the measurements of resistivity and Hall effect. As shown in Figs. 4(a)-4(d), in two cases where the magnetic field is along $\textbf{H} \parallel b$ and $\textbf{H} \perp bc$, the current is applied perpendicular and parallel to the distorted kagome plane, respectively. The resistivity and Hall resistivity are measured simultaneously. Figures 4(e)-4(h) show the measurement results of resistivity and Hall effect under increasing and decreasing fields for four configurations, as well as the comparison with the magnetization. In Figs. 4(e) and 4(f), both MR and Hall resistivity exhibit three transitions, corresponding to the three changes in spin states present in the magnetization, indicating a strong correlation between electronic transport and magnetism. Before and after the formation of the magnetization plateaus, spin disorder and order respectively enhance and weaken the magnetic scattering, resulting in a peak in the MR curves. When the magnetic field exceeds 7.3 T, the magnetization enters a saturation state, and the MR continuously decreases as the magnetic field increases, forming the negative MR values of $-$45\% ($\textbf{I} \parallel c$, $\textbf{H} \parallel b$) and $-$31\% ($\textbf{I} \perp bc$, $\textbf{H} \parallel b$) at 1.8 K and 14 T (Supplemental Material\cite{SupplementalMaterial}). The Hall resistivity $\rho_{xy}$ exhibits significant AHE at the 1/3 and 1/2 magnetization plateaus. In Figs. 4(g) and 4(h), since the magnetic field is along the $\textbf{H} \perp bc$ direction, the electronic transport properties become more complex. Between H$_3$ and H$_4$, the 2/3 magnetization appears as a broad peak in MR but does not induce the change in $\rho_{xy}$, likely due to the weak signal. The 5/6 magnetization exhibits a kink in the MR and Hall resistivity at H$_4$, after which the magnetization gradually increases to the saturation state. The transition to saturation magnetization at H$_5$ can also be observed in MR and Hall resistivity. The position of H$_5$ in Figs. 4(g) and 4(h) is different, which may be due to the gradual and slow nature of this transition rather than a sudden change. Since the magnetic moments are constrained within the distorted kagome plane, the application of in-plane current allows carriers to perceive the change in magnetic order at the onset of the transition.

One of the most striking features in Fig. 4 is the significant finite hysteresis of MR and Hall resistivity at the 1/2 magnetization plateau, while the hysteresis almost vanishes in the magnetic field dependent magnetization. Although the hysteresis of $\rho_{xy}$ in Fig. 4(f) is relatively small, it still exists. As the temperature increases, hysteresis in MR and Hall resistivity gradually weakens and disappears (Supplemental Material\cite{SupplementalMaterial}). These observations indicate the existence of time-reversal-like degenerate states in Ce$_3$MgBi$_5$, similar to that in HoAgGe, which are closely related to the reversal of toroidal moment states. The clockwise and counterclockwise toroidal moment states have opposite chirality and Berry curvature. The application of an in-plane magnetic field will reverse the chirality of the toroidal structures and alter the distribution of degenerate states, thereby affecting the MR and Hall effect but not the net magnetization. During the decrease of the magnetic field, these changes will be preserved rather than restored to the previous state due to the hindrance of the distorted kagome lattice. Therefore, hysteresis appears in MR and Hall effect but is absent in magnetization measurements.

%It is interesting to compare the hysteresis sizes in the four measurement configurations. When current is applied within the distorted kagome lattice plane [Figs. 4(f) and 4(h)], the hysteresis size of MR is significantly larger than that of Hall effect, indicating that measuring in-plane voltage is easier to distinguish the degenerate states than measuring out-of-plane voltage.

It is interesting to compare the experimental results of Ce$_3$MgBi$_5$ with those of HoAgGe. It is worth noting that hysteresis occurs at the 1/2 plateau in Ce$_3$MgBi$_5$, while it occurs at the 1/3 and 2/3 plateaus in HoAgGe (the 1/2 plateau is absent in HoAgGe). We further explored the underlying mechanism. Figure 5(a) shows the chirality reversal model of the toroidal moment states of HoAgGe at the 1/3 plateau, which is equivalent to the result of the superposition of $\mathcal{D}$ and $\emph{R}_b^{\pi}$ operations. Here, $\mathcal{D}$ is a special time-reversal-like operation that reverses the distortion of the lattice while maintaining the spin degree of freedom unchanged, and $\emph{R}_b^{\pi}$ is an operation that rotates both orbital and spin degrees of freedom by $\pi$ about the $b$ axis\cite{zhao2024discrete}. However, the stacking pattern of distorted kagome layers along the $c$ axis in Ce$_3$MgBi$_5$ and HoAgGe is different. The stacking in HoAgGe is an overlapping arrangement, while in Ce$_3$MgBi$_5$, adjacent layers are rotated by 180$^{\circ}$, forming two sets of sublattices [Figs. 1(b) and 1(c)]. Figures 5(b) and 5(c) respectively show the transformation processes of two sublattices with the same chirality in Ce$_3$MgBi$_5$ under $\mathcal{D}$ and $\emph{R}_b^{\pi}$ operations. It can be seen that the sublattice-1 after operation $\mathcal{D}$ is actually equivalent to the initial state of sublattice-2. Similarly, the sublattice-2 after operation $\mathcal{D}$ is also equivalent to the initial state of sublattice-1. The two sublattices with the same chirality did not cause any change in the system after operation $\mathcal{D}$, indicating that operation $\mathcal{D}$ is ineffective in this case. However, $\mathcal{D}$ is important for the change of MR and Hall resistivity, as it can change the Berry curvature of the occupied states\cite{zhao2024discrete}. Therefore, the absence of hysteresis at the 1/3 plateau in Ce$_3$MgBi$_5$ can be attributed to the inefficiency of the time-reversal-like operation $\mathcal{D}$ due to its special crystal structure. The 2/3 plateau in HoAgGe becomes more complex due to the involvement of $c$-axis spin order, but the transition between degenerate states is also related to the operation $\emph{R}_b^{\pi}\mathcal{D}$\cite{zhao2024discrete}. The absence of hysteresis at the 2/3 plateau in Ce$_3$MgBi$_5$ can also be understood in a similar way.

\begin{figure}[htbp]
\centering
\includegraphics[width=0.48\textwidth]{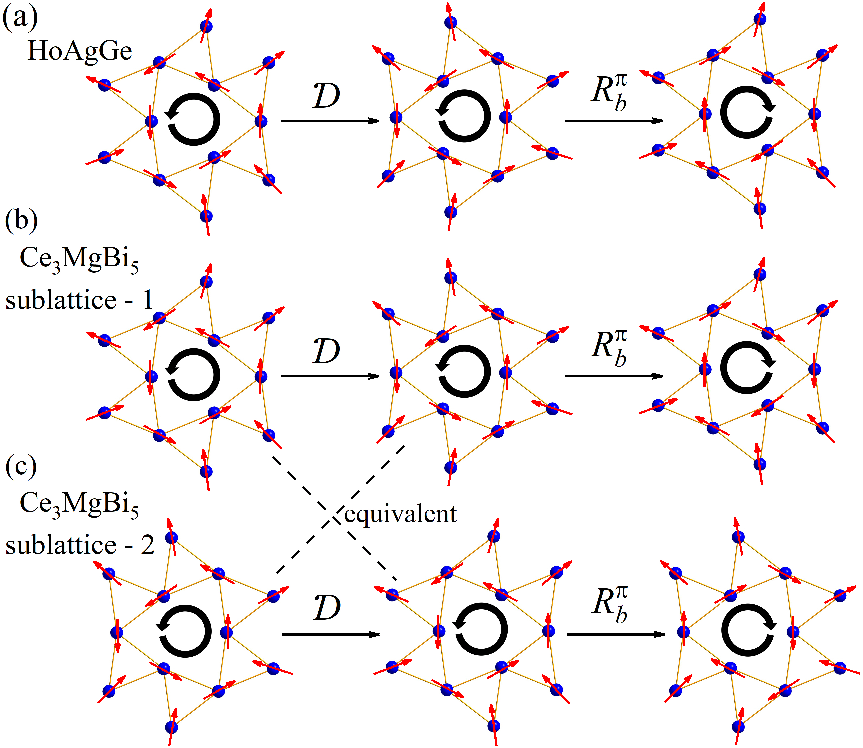}
\caption{(a) Model of the chirality reversal process of the toroidal moment states in HoAgGe under $\mathcal{D}$ and $\emph{R}_b^{\pi}$ operations. (b) and (c) The effects of $\mathcal{D}$ and $\emph{R}_b^{\pi}$ operations on the two sublattices of Ce$_3$MgBi$_5$.}
\end{figure}

It should be noted that the above analysis is based on the assumption that the two sublattices in Ce$_3$MgBi$_5$ have the same chirality. If the two sublattices have different chirality, the effectiveness of operation $\mathcal{D}$ will not be eliminated and hysteresis will emerge, which may be the case at the 1/2 plateau. With the increase of the magnetic field, the two sublattices evolve from the same chirality at the 1/3 plateau to the opposite chirality at the 1/2 plateau, resulting in the absence and appearance of hysteresis. Further neutron diffraction experiments are necessary to study the changes in the magnetic structure of Ce$_3$MgBi$_5$.

In summary, single crystals of a new compound Ce$_3$MgBi$_5$ are grown. The crystal structure, magnetic properties, and electrical transport properties have been systematically studied. Ce$_3$MgBi$_5$ exhibits complex AFM transitions and a series of fractional magnetization plateaus. During the rise and fall of the magnetic field, the significant hysteresis in MR and Hall resistivity at the 1/2 magnetization plateau and the vanishing hysteresis in magnetization indicate the existence of hidden time-reversal-like degenerate states in Ce$_3$MgBi$_5$, which is related to the chirality of the toroidal moment states. Further analysis suggests that the unique crystal structure of Ce$_3$MgBi$_5$ may have a shielding effect on the time-reversal-like operation. Our study will be beneficial for designing or exploring similar degenerate states in other materials.

\emph{Acknowledgments.} This work is supported by the National Key Research and Development Program of China (Grant No. 2023YFA1406500) and the National Natural Science Foundation of China (Grants No. 12474098, No. 12574042, No. 12274388, No. 12404043, and No. 12174361), and the Natural Science Foundation of Anhui Province (Grant No. 2408085QA024).

\bibliography{Bibtex}
\end{document}

\title{Supplementary Information for: The distinction of time-reversal-like degeneracy by electronic transport in a new compound Ce$_3$MgBi$_5$}

%% Notice placement of commas and superscripts and use of &
%% in the author list
\author{Yi-Yan Wang}%\email{wyy@ahu.edu.cn}
\affiliation{Anhui Provincial Key Laboratory of Magnetic Functional Materials and Devices, Institutes of Physical Science and Information Technology, Anhui University, Hefei, Anhui 230601, China}
\affiliation{Key Laboratory of Materials Physics, Ministry of Education, School of Physics, Zhengzhou University, Zhengzhou 450001, China}

\author{Ping Su}
\affiliation{Anhui Provincial Key Laboratory of Magnetic Functional Materials and Devices, Institutes of Physical Science and Information Technology, Anhui University, Hefei, Anhui 230601, China}

\author{Kai-Yuan Hu}
\affiliation{Anhui Provincial Key Laboratory of Magnetic Functional Materials and Devices, Institutes of Physical Science and Information Technology, Anhui University, Hefei, Anhui 230601, China}

\author{Yi-Ran Li}
\affiliation{Anhui Provincial Key Laboratory of Magnetic Functional Materials and Devices, Institutes of Physical Science and Information Technology, Anhui University, Hefei, Anhui 230601, China}

\author{Na Li}
\affiliation{Anhui Provincial Key Laboratory of Magnetic Functional Materials and Devices, Institutes of Physical Science and Information Technology, Anhui University, Hefei, Anhui 230601, China}

\author{Ying Zhou}
\affiliation{Anhui Provincial Key Laboratory of Magnetic Functional Materials and Devices, Institutes of Physical Science and Information Technology, Anhui University, Hefei, Anhui 230601, China}

\author{Dan-Dan Wu}
\affiliation{Anhui Provincial Key Laboratory of Magnetic Functional Materials and Devices, Institutes of Physical Science and Information Technology, Anhui University, Hefei, Anhui 230601, China}

\author{Yan Sun}
\affiliation{Anhui Provincial Key Laboratory of Magnetic Functional Materials and Devices, Institutes of Physical Science and Information Technology, Anhui University, Hefei, Anhui 230601, China}

\author{Qiu-Ju Li}
\affiliation{School of Physics, Anhui University, Hefei, Anhui 230601, China}

\author{Xia Zhao}
\affiliation{School of Physical Sciences, University of Science and Technology of China, Hefei, Anhui 230026, China}

\author{Hui Liang}%\email{lianghui@ahu.edu.cn}
\affiliation{Anhui Provincial Key Laboratory of Magnetic Functional Materials and Devices, Institutes of Physical Science and Information Technology, Anhui University, Hefei, Anhui 230601, China}

\author{Xue-Feng Sun}%\email{xfsun@ahu.edu.cn}
\affiliation{Anhui Provincial Key Laboratory of Magnetic Functional Materials and Devices, Institutes of Physical Science and Information Technology, Anhui University, Hefei, Anhui 230601, China}
\maketitle
%\newpage

\section{Experimental methods}

Single crystals of Ce$_3$MgBi$_5$ were grown by the flux method. The atomic proportions have been checked by energy dispersive x-ray spectroscopy (EDS). The crystal structure of Ce$_3$MgBi$_5$ was determined by the single crystal x-ray diffraction experiment, which was performed using a Bruker D8 Quest diffractometer (Mo K$_{\alpha}$). The measurements of resistivity and Hall effect were performed on a Quantum Design physical property measurement system (PPMS-14T). The magnetic properties were measured on a Quantum Design magnetic property measurement system (MPMS-7T). The magnetization up to 14 T was measured with a vibrating sample magnetometer (VSM) option of PPMS.
%The starting materials Ce, Mg, and Bi were placed into an alumina crucible with a molar ratio of Ce : Mg : Bi = 1 : 6 : 9. The crucible was then sealed in a quartz tube under high-vacuum condition. After heating to 900 $^\circ$C and holding for 20 h, the quartz tube was cooled to 700 $^\circ$C at a rate of 2 $^\circ$C/h, at which the excess flux was removed using a centrifuge.

\begin{table}
\setlength{\tabcolsep}{0pt}
  \renewcommand{\arraystretch}{1.4}
  \centering
  \caption{Crystallographic parameters derived from single-crystal x-ray data of Ce$_3$MgBi$_5$ at 296.15 K.}
  \begin{tabular}{cccccc}
    \toprule
          & Space group & P6$_3$/mcm (No.193) & \\
          & Crystal system & hexagonal & \\
          & Formula weight & 1489.57 & \\
          & Unit cell dimensions & $a=b=$ 9.638(5)${\AA}$, $c=$ 6.520(5)${\AA}$ & \\
          &  & $\alpha=\beta=$ 90$^\circ$, $\gamma=$ 120$^\circ$ & \\
          & Volume (${\AA}^3$) & 524.5(7) & \\
          & Calcd. density (g/cm$^3$ ) & 9.432 & \\
          & Wavelength (${\AA}$) & 0.71073 & \\
          & Z & 2 & \\
          & Abs. coeff. (mm$^{-1}$) & 96.267 & \\
          & F(000) & 1202.0 & \\
          & 2$\theta$ range (deg) & 4.88 -- 66.478 & \\
          & Ranges in \emph{hkl} & $-$14$\leqslant$\emph{h}$\leqslant$7, $-$14$\leqslant$\emph{k}$\leqslant$14, $-$9$\leqslant$\emph{l}$\leqslant$5 & \\
          & Refls. collected & 4893 & \\
          & Ind. refls. & 393 [R(int) $=$ 0.0912] & \\
          & GOF on F$^2$ & 1.125 & \\
          & Final R indices [I $>$ 2$\sigma$(I)] & R1 $=$ 0.0741, wR2 $=$ 0.1771 & \\
          & Final R indices (all data) & R1 $=$ 0.0843, wR2 $=$ 0.1874 & \\
          & Largest diff. peak/hole (e${\AA}^{-3}$) & 7.71/$-$5.00 & \\
  \bottomrule
  \end{tabular}
\end{table}

\begin{figure*}[htbp]
\centering
\includegraphics[width=0.6\textwidth]{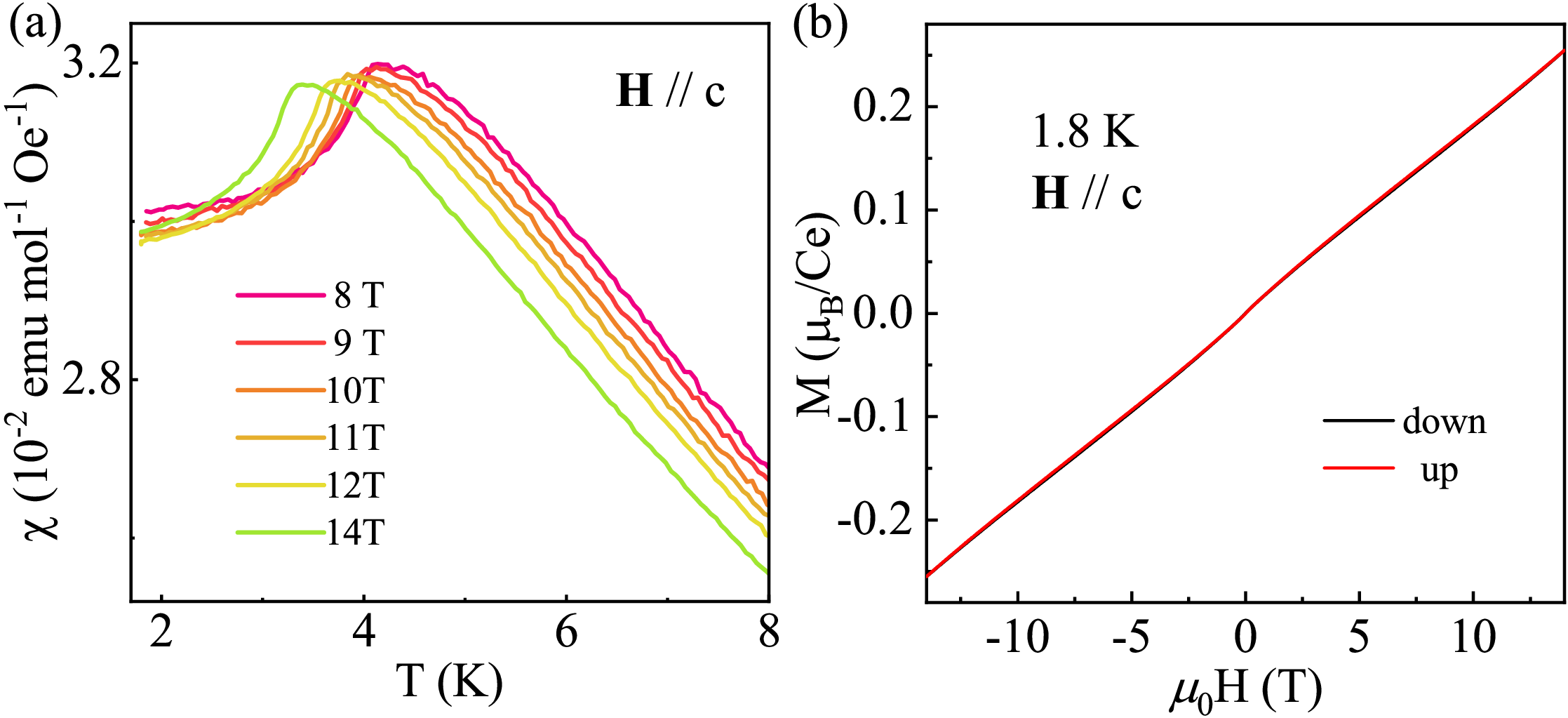}
\caption{Temperature dependent magnetic susceptibility (8-14 T) and magnetic field dependent magnetization in the direction of $\textbf{H} \parallel c$.}
\end{figure*}

\begin{figure*}[htbp]
\centering
\includegraphics[width=0.3\textwidth]{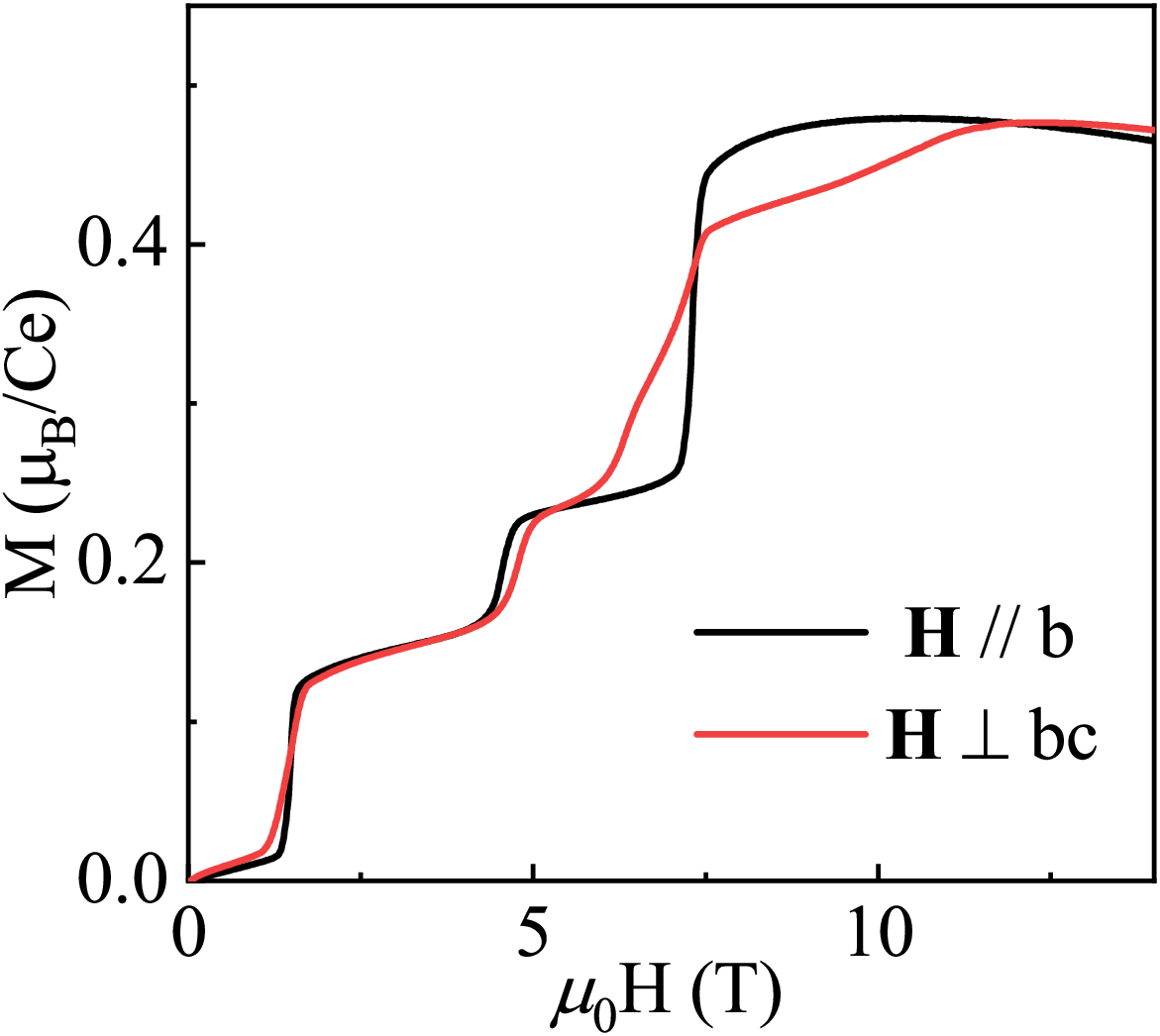}
\caption{Comparison of magnetization curves in the $\textbf{H} \perp bc$ and $\textbf{H} \parallel b$ directions at 1.8 K after subtracting the Van-Vleck paramagnetic background.}
\end{figure*}

\begin{figure*}[htbp]
\centering
\includegraphics[width=0.9\textwidth]{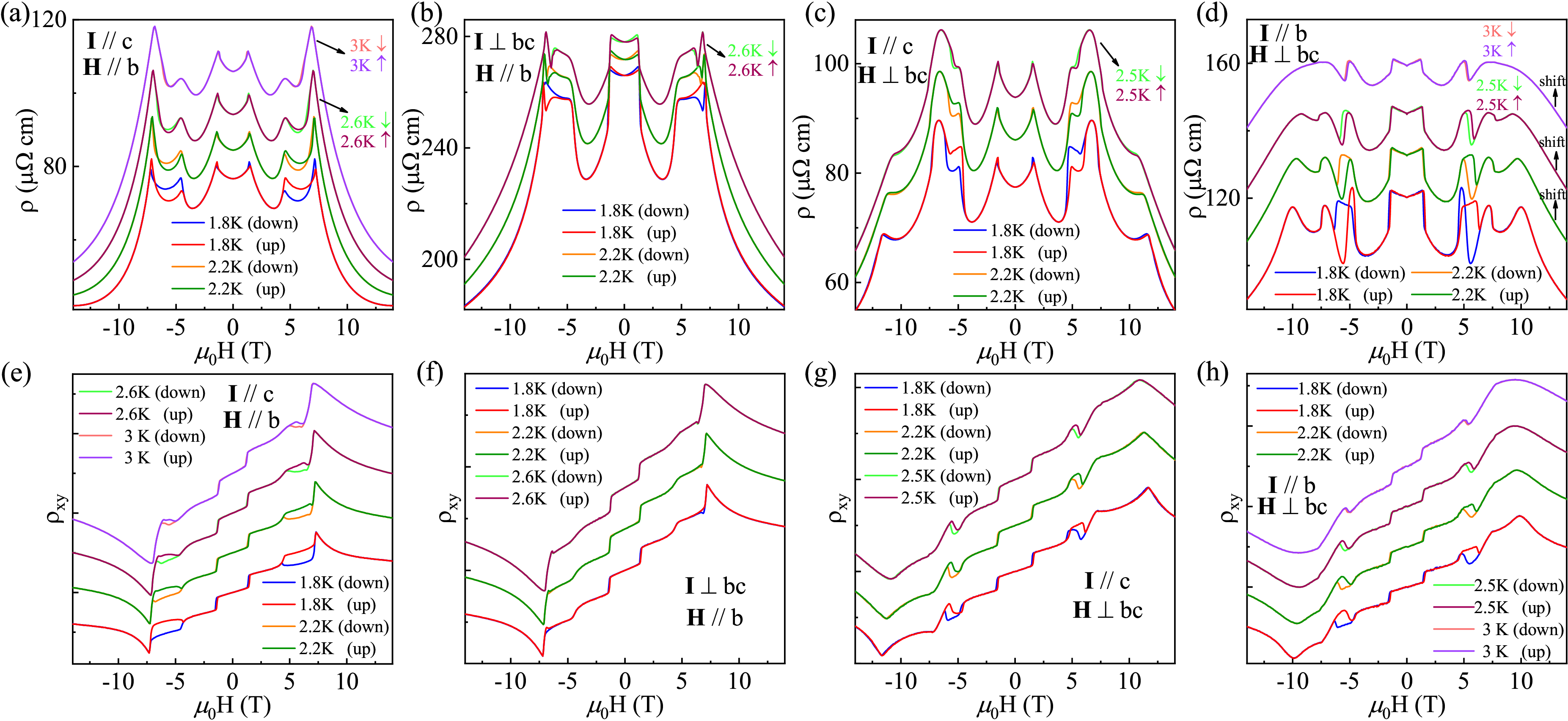}
\caption{(a)-(d) Hysteresis of MR during the rise and fall of the magnetic field at different temperatures. (e)-(h) Hysteresis of Hall resistivity during the rise and fall of the magnetic field at different temperatures. In (d)-(h), the curves have been shifted along the vertical axis for clearer display. The measurement configurations correspond to Figs. 4(a)-4(d) in the main text, respectively.}
\end{figure*}

\begin{figure*}[htbp]
\centering
\includegraphics[width=0.9\textwidth]{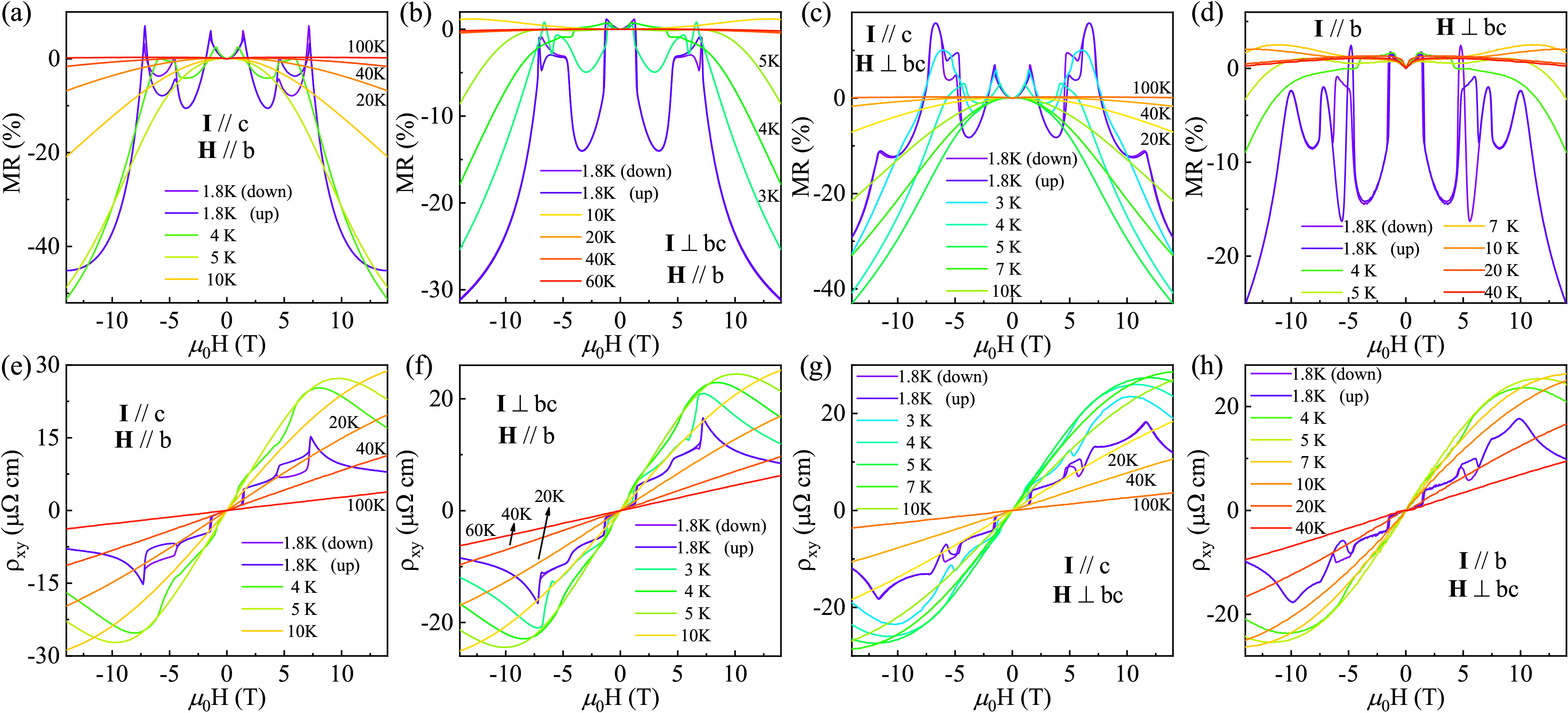}
\caption{(a)-(d) Magnetic field dependent MR at different temperatures. (e)-(h) Magnetic field dependent Hall resistivity at different temperatures. The measurement configurations correspond to Figs. 4(a)-4(d) in the main text, respectively.}
\end{figure*}

\begin{figure*}[htbp]
\centering
\includegraphics[width=0.9\textwidth]{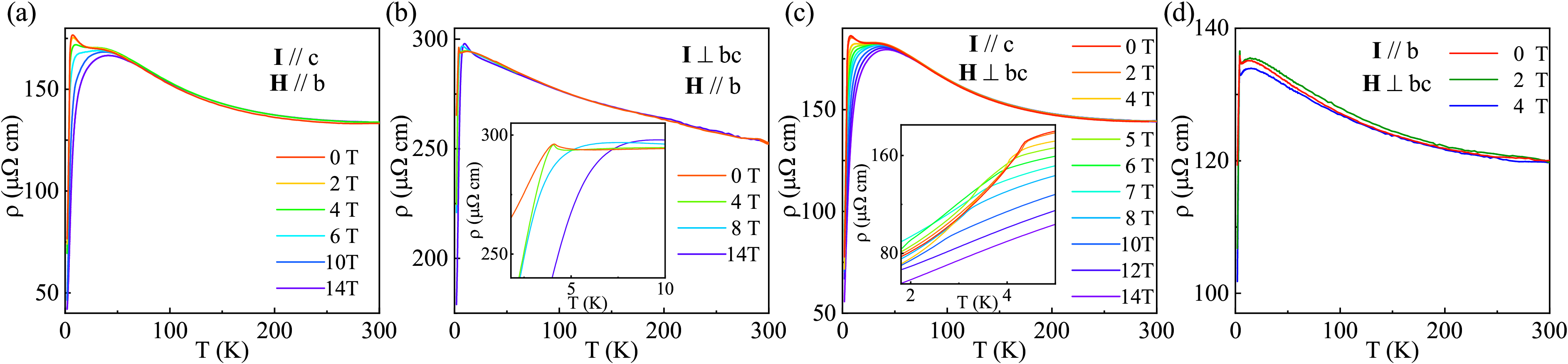}
\caption{(a)-(d) Temperature dependent resistivity at different magnetic fields. The measurement configurations correspond to Figs. 4(a)-4(d) in the main text, respectively. Insets: the enlarged part at low temperatures.}
\end{figure*}